\documentclass[12pt,a4paper]{article}
\usepackage{amsmath}
\usepackage{amssymb}
\usepackage{amscd}
\usepackage{latexsym}
\usepackage{graphicx,color}
\makeatletter
\def\rddots{\mathinner{\mkern1mu\raise\p@%
    \vbox{\kern7\p@\hbox{.}}\mkern2mu%
    \raise4\p@\hbox{.}\mkern2mu\raise7\p@\hbox{.}\mkern1mu}}
\makeatother
\newcommand{\eqdef}{\stackrel{\text{def}}{=}}

\newcommand{\ket}[1]{{\vert{#1}\rangle}}

\newcommand{\fukuso}{{\mathbf C}}
\newcommand{\real}{{\mathbf R}}
\newcommand{\futon}{{\bf N}}

\newcommand{\tr}{{\rm tr}}

\begin{document}

\title{\sl SO(4) Re-revisited}
\author{
  Kazuyuki FUJII
  \thanks{E-mail address : fujii@yokohama-cu.ac.jp }\\
  Department of Mathematical Sciences\\
  Yokohama City University\\
  Yokohama, 236--0027\\
  Japan
  }
\date{}
\maketitle
\begin{center}
\begin{Large}
To the memory of Steve Jobs \footnote{Steven 
Paul Jobs (1955--2011)}
\end{Large}
\end{center}
\vspace{5mm}
\begin{abstract}
  In this note an explicit expression of $\exp A\ (A\in so(4))$ is given 
  in terms of the magic matrix by Makhlin.
\end{abstract}
%
%

%
%
%     Honbun
%
%
\section{Introduction}

The four--dimensional special orthogonal group $SO(4)$ plays 
an important role in both Mathematics and Physics, and maybe 
in Chemistry. Since it is semi--simple, essential properties 
reduce to those of $SU(2)$ or $SO(3)\ (\cong SU(2)/{\bf Z}_{2})$. 
See \cite{Five} for a comprehensive introduction to this topic and we recommend \cite{IY} 
as a good text-book of  Group and Topology.

Let us recall the definition:
\begin{equation}
SO(4)=\{O\in M(4;\real)\ |\ O^{t}O=OO^{t}=I_4,\ \det O=1\},
\end{equation}
where $t$ denotes the transpose, $I_4$ the four dimensional unit matrix and 
det\,$O$ the determinant of $O$.  Its Lie algebra is given by
\begin{equation}
so(4)=\{A\in M(4;\real)\ |\ A^{t}=-A,\ \tr A=0\}.
\end{equation}
Then it is well--known that every element of $SO(4)$ can be 
written as
\begin{equation}
SO(4)=\{e^{A}\ |\ A\in so(4)\}
\end{equation}
where $e^{A}\ (=\exp A)$ is the exponential defined by
\begin{equation}
e^{A}=I_4+A+\frac{1}{2!}A^{2}+\cdots +\frac{1}{n!}A^{n}+\cdots .
\end{equation}

The form is simple and beautiful, while to calculate $e^{A}$  is 
another problem. In fact, it is very difficult and 
its explicit form has not been reported as far as we know. 
In this note we re-revisit this problem and give a ``super smart" 
form (see the concluding remarks) to $e^{A}$ in terms of 
the magic matrix by Makhlin.

Let us set up the problem once more. \ Calculate the exponential
\begin{equation}
e^{A}\ ; \ 
A=
\left(
  \begin{array}{cccc}
     0      & f_{12}  & f_{13}   & f_{14}   \\
   -f_{12} & 0       & f_{23}   & f_{24}   \\
   -f_{13} & -f_{23} & 0        & f_{34}   \\
   -f_{14} & -f_{24} & -f_{34} & 0 
  \end{array}
\right)\ \in\ so(4)
\end{equation}
{\bf explicitly}. For this purpose we need the magic matrix 
by Makhlin \cite{YMa}.

\section{Magic Matrix}

In this section we review the result in \cite{FOS}, \cite{FS} 
within our necessity, which is a bit different from the original in \cite{YMa}. 
This section is also a brief introduction to {\bf Quantum Computation} 
for undergraduates.

The 1--qubit space is the two dimensional vector space over $\fukuso$, 
namely
\[
\fukuso^{2}
=\mbox{Vect}_{\fukuso}\{\ket{0},\ket{1}\}
\equiv \{\alpha\ket{0}+\beta\ket{1}\ |\ \alpha, \beta \in \fukuso\},
\]
where
\begin{equation}
\label{eq:bra-ket}
\ket{0}=
\left(
\begin{array}{c}
 1 \\
 0
\end{array}
\right),
\quad 
\ket{1}=
\left(
\begin{array}{c}
 0 \\
 1
\end{array}
\right).
\end{equation}
Let $\{\sigma_{1}, \sigma_{2}, \sigma_{3}\}$ be the Pauli matrices acting on 
the space
\begin{equation}
\label{eq:Pauli matrices}
\sigma_{1} = 
\left(
  \begin{array}{cc}
    0 & 1 \\
    1 & 0
  \end{array}
\right), \quad 
\sigma_{2} = 
\left(
  \begin{array}{cc}
    0 & -i \\
    i & 0
  \end{array}
\right), \quad 
\sigma_{3} = 
\left(
  \begin{array}{cc}
    1 & 0 \\
    0 & -1
  \end{array}
\right)
\end{equation}
and we denote by $1_{2}$ the two dimensional unit matrix
\begin{equation}
\label{eq:nuit matrix}
1_{2}=
\left(
  \begin{array}{cc}
    1 & 0 \\
    0 & 1
  \end{array}
\right).
\end{equation}

Next let us consider the 2--qubit space. 
%Now we use notations on tensor product which are different from usual ones. 
%That is, 
%
We start with the Kronecker product $a\otimes b$ of two $\fukuso^{2}$ 
vectors $a$ and $b$:
\[
a\otimes b
\equiv
\left(
  \begin{array}{c}
    a_{1} \\
    a_{2}
  \end{array}
\right)
\otimes
\left(
  \begin{array}{c}
    b_{1} \\
    b_{2}
  \end{array}
\right)
\equiv 
\left(
  \begin{array}{c}
    a_{1}b_{1} \\
    a_{1}b_{2} \\
    a_{2}b_{1} \\
    a_{2}b_{2}
  \end{array}
\right).
\]
They form a set of tensor products:
\[
\fukuso^{2}{\otimes}\fukuso^{2}=\{a\otimes b\ |\ a,b\in \fukuso^{2}\}.
\]
The 2--qubit space is a vector space generated by them: %
\[
\fukuso^{2}\widehat{\otimes}\fukuso^{2}=
\left\{\sum_{j=1}^{k}\lambda_{j}a_{j}\otimes b_{j}\ |\ a_{j},b_{j}\in \fukuso^{2},
\ \lambda_{j}\in \fukuso,\ k\in \futon \right\}\cong \fukuso^{4}.
\]
Then we have
\[
\fukuso^{2}\widehat{\otimes}\fukuso^{2}=\mbox{Vect}_{\fukuso}
\{\ket{00},\ket{01},\ket{10},\ket{11}\}
\]
where $\ket{ij}=\ket{i}\otimes \ket{j}\ (i, j\in \{0,1\})$ for simplicity. 

For $A,\ B\in M(2;\fukuso)$
\[
A=\left(
  \begin{array}{cc}
    a_{11} & a_{12} \\
    a_{21} & a_{22} 
  \end{array}
\right),\quad
B=\left(
  \begin{array}{cc}
    b_{11} & b_{12} \\
    b_{21} & b_{22} 
  \end{array}
\right)
\]
the Kronecker product $A\otimes B$ is defined by
\[
(A\otimes B)(a\otimes b)=(Aa)\otimes (Bb).
\]
Its explicit form is given by
\begin{eqnarray*}
A\otimes B
&=&
\left(
  \begin{array}{cc}
    a_{11} & a_{12} \\
    a_{21} & a_{22} 
  \end{array}
\right)
\otimes
\left(
  \begin{array}{cc}
    b_{11} & b_{12} \\
    b_{21} & b_{22} 
  \end{array}
\right)  \\
&=&
\left(
  \begin{array}{cccc}
    a_{11}b_{11} & a_{11}b_{12} & a_{12}b_{11} & a_{12}b_{12}  \\
    a_{11}b_{21} & a_{11}b_{22} & a_{12}b_{21} & a_{12}b_{22}  \\
    a_{21}b_{11} & a_{21}b_{12} & a_{22}b_{11} & a_{22}b_{12}  \\
    a_{21}b_{21} & a_{21}b_{22} & a_{22}b_{21} & a_{22}b_{22}    
  \end{array}
\right).
\end{eqnarray*}
Readers should check it. From this we have the multiplication
\[
(A_{1}\otimes B_{1})(A_{2}\otimes B_{2})=A_{1}A_{2}\otimes B_{1}B_{2}
\]
for $A_{1},A_{2},B_{1},B_{2}\in M(2;\fukuso)$.

By $H_{0}(2;\fukuso)$ we denote the set of all traceless hermitian matrices in 
$M(2;\fukuso)$. Then it is well--known
\[
H_{0}(2;\fukuso)=\{a\equiv a_{1}\sigma_{1}+a_{2}\sigma_{2}+a_{3}\sigma_{3}\ 
|\ a_{1},a_{2},a_{3}\in \real\}
\]
and $H_{0}(2;\fukuso)\cong su(2)$ where $su(2)$ is the Lie algebra of the 
group $SU(2)$
\[
su(2)=\{i\left(a_{1}\sigma_{1}+a_{2}\sigma_{2}+a_{3}\sigma_{3}\right)\ 
|\ a_{1},a_{2},a_{3}\in \real\}.
\]

\vspace{3mm}
The (famous) Bell bases 
$\{\ket{\Psi_{1}},\ket{\Psi_{2}},\ket{\Psi_{3}},\ket{\Psi_{4}}\}$ 
in $\fukuso^{2}\widehat{\otimes}\fukuso^{2}$ are defined by
\begin{eqnarray}
\label{eq:Bell bases}
\ket{\Psi_{1}}&=&\frac{1}{\sqrt{2}}(\ket{00}+\ket{11}),\quad
\ket{\Psi_{2}}=\frac{1}{\sqrt{2}}(\ket{01}+\ket{10}), \nonumber \\
\ket{\Psi_{3}}&=&\frac{1}{\sqrt{2}}(\ket{01}-\ket{10}),\quad
\ket{\Psi_{4}}=\frac{1}{\sqrt{2}}(\ket{00}-\ket{11}),
\end{eqnarray}
and by making use of them we can give 
the isomorphism ($SU(2)\otimes SU(2) \cong SO(4)$) as 
the adjoint action (the Makhlin's theorem) as follows
\[
F : SU(2)\otimes SU(2) \longrightarrow SO(4),\quad 
F(P\otimes Q)=R^{\dagger}(P\otimes Q)R
\]
where
\begin{equation}
R=
\left(
\ket{\Psi_{1}},-i\ket{\Psi_{2}},-\ket{\Psi_{3}},-i\ket{\Psi_{4}}
\right)
=\frac{1}{\sqrt{2}}
\left(
  \begin{array}{cccc}
    1 &  0 &  0 & -i  \\
    0 & -i & -1 &  0  \\
    0 & -i &  1 &  0  \\
    1 &  0 &  0 &  i 
  \end{array}
\right).
\end{equation}
Note that the unitary matrix $R$ is a bit different from $Q$ in \cite{YMa}.

\vspace{3mm}
Let us consider this problem at the Lie algebra level because it is in general 
not easy to treat it directly at the  Lie group level:
%\vspace{5mm}
%%
%\begin{center}
%\input{Lie-diagram.fig}
%\end{center}
%
\begin{align*}
\begin{CD}
\mathfrak{L}(SU(2)\otimes SU(2))\ 
\quad @> f>> \quad
\mathfrak{L}(SO(4))\ \\[4pt]
@V\mbox{exp}\ \ VV
@VV\ \ \mbox{exp}V\\[4pt]
SU(2)\otimes SU(2)
\quad @>>F> \quad
SO(4)
\end{CD} \qquad .
\end{align*}

\vspace{5mm} \noindent
Since the Lie algebra of $SU(2)\otimes SU(2)$ is
\[
\mathfrak{L}(SU(2)\otimes SU(2))=
\left\{i(a\otimes 1_{2}+1_{2}\otimes b)\ |\ a,b \in H_{0}(2;\fukuso)\right\},
\]
we have only to examine
\begin{equation}
\label{eq:Lie algebra level}
f(i(a\otimes 1_{2}+1_{2}\otimes b))=
iR^{\dagger}(a\otimes 1_{2}+1_{2}\otimes b)R\in 
\mathfrak{L}(SO(4))\equiv so(4).
\end{equation}
For $a=\sum_{j=1}^{3}a_{j}\sigma_{j}$ and $b=\sum_{j=1}^{3}b_{j}\sigma_{j}$ 
some algebra gives
\begin{equation}
\label{eq:correspondence}
iR^{\dagger}(a\otimes 1_{2}+1_{2}\otimes b)R 
=
\left(
  \begin{array}{cccc}
    0 & a_{1}+b_{1} & a_{2}-b_{2} & a_{3}+b_{3}       \\
    -(a_{1}+b_{1}) & 0 & a_{3}-b_{3} & -(a_{2}+b_{2}) \\
    -(a_{2}-b_{2}) & -(a_{3}-b_{3}) & 0 & a_{1}-b_{1} \\
    -(a_{3}+b_{3}) & a_{2}+b_{2} & -(a_{1}-b_{1}) & 0  
  \end{array}
\right).
\end{equation}
Conversely, for
\[
A=
\left(
  \begin{array}{cccc}
    0 & f_{12} & f_{13} & f_{14}   \\
   -f_{12} & 0 & f_{23} & f_{24}   \\
   -f_{13} & -f_{23} & 0 & f_{34}  \\
   -f_{14} & -f_{24} & -f_{34} & 0  
  \end{array}
\right)\in so(4)
\]
some algebra gives
\begin{equation}
\label{eq:adjoint}
RAR^{\dagger}=i(a\otimes 1_{2}+1_{2}\otimes b)
\end{equation}
with
\begin{eqnarray}
\label{eq:left-a}
a=a_{1}\sigma_{1}+a_{2}\sigma_{2}+a_{3}\sigma_{3}
   &=&\frac{f_{12}+f_{34}}{2}\sigma_{1}+
    \frac{f_{13}-f_{24}}{2}\sigma_{2}+
    \frac{f_{14}+f_{23}}{2}\sigma_{3}, \\
\label{eq:right-b}
b=b_{1}\sigma_{1}+b_{2}\sigma_{2}+b_{3}\sigma_{3}
   &=&\frac{f_{12}-f_{34}}{2}\sigma_{1}-
    \frac{f_{13}+f_{24}}{2}\sigma_{2}+
    \frac{f_{14}-f_{23}}{2}\sigma_{3},\\
    a_{1}&=&\frac{f_{12}+f_{34}}{2},\ \ a_{2}=\frac{f_{13}-f_{24}}{2},\ \ \ \ a_{3}=\frac{f_{14}+f_{23}}{2}, \nonumber\\
b_{1}&=&\frac{f_{12}-f_{34}}{2},\ \ b_{2}=-\frac{f_{13}+f_{24}}{2},\ \ b_{3}=\frac{f_{14}-f_{23}}{2}.\nonumber
\end{eqnarray}
Readers should check it.

\vspace{3mm}\noindent
{\bf Comment}\ \ 
It is very interesting to note that $a$ is the {\bf self--dual} part 
and $b$ the {\bf anti--self--dual} one. See \cite{FOS-2} for details.

\vspace{3mm}
Last, we list a well--known formula for the exponentiation of $su(2)$ for later convenience. For
\[
X=x_{1}\sigma_{1}+x_{2}\sigma_{2}+x_{3}\sigma_{3},\quad 
|X|\eqdef \sqrt{x_{1}^{2}+x_{2}^{2}+x_{3}^{2}},\quad x_j\in \mathbf{R},
\]
it is easy to obtain the following
\begin{eqnarray}
\label{eq:fundamental formula}
e^{iX}
&=&\cos|X|1_{2}+\frac{\sin|X|}{|X|} iX \nonumber \\
&=&
\left(
  \begin{array}{cc}
\displaystyle    \cos|X|+\frac{\sin|X|}{|X|}ix_{3} & \displaystyle \frac{\sin|X|}{|X|}i(x_{1}-ix_{2}) \\
 \displaystyle   \frac{\sin|X|}{|X|}i(x_{1}+ix_{2}) &  \displaystyle \cos|X|-\frac{\sin|X|}{|X|}ix_{3}
  \end{array}
\right).
\end{eqnarray}
\vspace{3mm}\noindent
From this and (\ref{eq:left-a}), (\ref{eq:right-b}) we have
\begin{eqnarray}
\label{eq:fundamental formula-a}
e^{ia}&=&
\left(
  \begin{array}{cc}
\displaystyle    \cos|a|+\frac{\sin|a|}{|a|}ia_{3} &  \displaystyle \frac{\sin|a|}{|a|}i(a_{1}-ia_{2}) \\
\displaystyle    \frac{\sin|a|}{|a|}i(a_{1}+ia_{2}) & \displaystyle \cos|a|-\frac{\sin|a|}{|a|}ia_{3}
  \end{array}
\right), \\[4pt]
\label{eq:fundamental formula-b}
e^{ib}&=&
\left(
  \begin{array}{cc}
\displaystyle    \cos|b|+\frac{\sin|b|}{|b|}ib_{3} & \displaystyle \frac{\sin|b|}{|b|}i(b_{1}-ib_{2}) \\
\displaystyle    \frac{\sin|b|}{|b|}i(b_{1}+ib_{2}) & \displaystyle \cos|b|-\frac{\sin|b|}{|b|}ib_{3}
  \end{array}
\right),
\end{eqnarray}
where
\[
|a|\eqdef\sqrt{a_{1}^{2}+a_{2}^{2}+a_{3}^{2}},\quad 
|b|\eqdef\sqrt{b_{1}^{2}+b_{2}^{2}+b_{3}^{2}}.
\]
%
%and
%
%\begin{eqnarray*}
%a_{1}&=&\frac{f_{12}+f_{34}}{2},\ \ a_{2}=\frac{f_{13}-f_{24}}{2},\ \ a_{3}=\frac{f_{14}+f_{23}}{2}, \\
%%
%b_{1}&=&\frac{f_{12}-f_{34}}{2},\ \ b_{2}=-\frac{f_{13}+f_{24}}{2},\ \ b_{3}=\frac{f_{14}-f_{23}}{2}.
%\end{eqnarray*}
%

\section{Exact Calculation of the Exponential}

The purpose of this section is to calculate\ \ $e^{A}$\ \ for 
\[
A=
\left(
  \begin{array}{cccc}
    0 & f_{12} & f_{13} & f_{14}   \\
   -f_{12} & 0 & f_{23} & f_{24}   \\
   -f_{13} & -f_{23} & 0 & f_{34}  \\
   -f_{14} & -f_{24} & -f_{34} & 0  
  \end{array}
\right)\in so(4).
\]

First, we have
\begin{eqnarray}
\label{eq:First}
e^{A}
&=&I_4\,e^{A}I_4 \nonumber \\
&=&R^{-1}Re^{A}R^{-1}R \ \ (\Longleftarrow R^{\dagger}=R^{-1}) \nonumber \\
&=&R^{-1}e^{RAR^{-1}}R \ \ (\Longleftarrow (\ref{eq:adjoint})) \nonumber \\
&=&R^{\dagger}e^{i(a\otimes 1_{2}+1_{2}\otimes b)}R \nonumber \\
&=&R^{\dagger}e^{i(a\otimes 1_{2})}e^{i(1_{2}\otimes b)}R \nonumber \\
&=&R^{\dagger}\left(e^{ia}\otimes 1_{2}\right)
              \left(1_{2}\otimes e^{ib}\right)R \nonumber \\
&=&R^{\dagger}\left(e^{ia}\otimes e^{ib}\right)R.
\end{eqnarray}

Second, in order to write down $e^{ia}\otimes e^{ib}$ we set
\begin{equation}
\label{eq:Second}
T\equiv e^{ia}\otimes e^{ib}=(t_{ij}).
\end{equation}
Then (\ref{eq:fundamental formula-a}) and 
(\ref{eq:fundamental formula-b}) give
\begin{eqnarray*}
t_{11}&=&
\left(\cos|a|+\frac{\sin|a|}{|a|}ia_{3}\right)
\left(\cos|b|+\frac{\sin|b|}{|b|}ib_{3}\right), \\
t_{21}&=&
\left(\cos|a|+\frac{\sin|a|}{|a|}ia_{3}\right)
\frac{\sin|b|}{|b|}i(b_{1}+ib_{2}), \\
t_{31}&=&
\frac{\sin|a|}{|a|}i(a_{1}+ia_{2})
\left(\cos|b|+\frac{\sin|b|}{|b|}ib_{3}\right), \\
t_{41}&=&
\frac{\sin|a|}{|a|}i(a_{1}+ia_{2})\frac{\sin|b|}{|b|}i(b_{1}+ib_{2}); 
\end{eqnarray*}
\begin{eqnarray*}
t_{12}&=&
\left(\cos|a|+\frac{\sin|a|}{|a|}ia_{3}\right)
\frac{\sin|b|}{|b|}i(b_{1}-ib_{2}), \\
t_{22}&=&
\left(\cos|a|+\frac{\sin|a|}{|a|}ia_{3}\right)
\left(\cos|b|-\frac{\sin|b|}{|b|}ib_{3}\right), \\
t_{32}&=&
\frac{\sin|a|}{|a|}i(a_{1}+ia_{2})\frac{\sin|b|}{|b|}i(b_{1}-ib_{2}), \\
t_{42}&=&
\frac{\sin|a|}{|a|}i(a_{1}+ia_{2})
\left(\cos|b|-\frac{\sin|b|}{|b|}ib_{3}\right); 
\end{eqnarray*}
\begin{eqnarray*}
t_{13}&=&
\frac{\sin|a|}{|a|}i(a_{1}-ia_{2})
\left(\cos|b|+\frac{\sin|b|}{|b|}ib_{3}\right), \\
t_{23}&=&
\frac{\sin|a|}{|a|}i(a_{1}-ia_{2})\frac{\sin|b|}{|b|}i(b_{1}+ib_{2}), \\
t_{33}&=&
\left(\cos|a|-\frac{\sin|a|}{|a|}ia_{3}\right)
\left(\cos|b|+\frac{\sin|b|}{|b|}ib_{3}\right), \\
t_{43}&=&
\left(\cos|a|-\frac{\sin|a|}{|a|}ia_{3}\right)
\frac{\sin|b|}{|b|}i(b_{1}+ib_{2}); 
\end{eqnarray*}
\begin{eqnarray*}
t_{14}&=&
\frac{\sin|a|}{|a|}i(a_{1}-ia_{2})\frac{\sin|b|}{|b|}i(b_{1}-ib_{2}), \\
t_{24}&=&
\frac{\sin|a|}{|a|}i(a_{1}-ia_{2})
\left(\cos|b|-\frac{\sin|b|}{|b|}ib_{3}\right), \\
t_{34}&=&
\left(\cos|a|-\frac{\sin|a|}{|a|}ia_{3}\right)
\frac{\sin|b|}{|b|}i(b_{1}-ib_{2}), \\
t_{44}&=&
\left(\cos|a|-\frac{\sin|a|}{|a|}ia_{3}\right)
\left(\cos|b|-\frac{\sin|b|}{|b|}ib_{3}\right).
\end{eqnarray*}

\vspace{3mm}
Third, let us write down $R^{\dagger}TR$. Some algebra gives
\begin{footnotesize}
\begin{eqnarray*}
\label{eq:Third}
&&R^{\dagger}TR \\
&=&\frac{1}{2}
\left(
  \begin{array}{cccc}
    1 & 0  & 0 & 1  \\
    0 & i   & i &  0  \\
    0 & -1 & 1 & 0  \\
    i  & 0  & 0 & -i 
  \end{array}
\right)
\left(
  \begin{array}{cccc}
    t_{11} & t_{12} & t_{13} & t_{14}  \\
    t_{21} & t_{22} & t_{23} & t_{24}  \\
    t_{31} & t_{32} & t_{33} & t_{34}  \\
    t_{41} & t_{42} & t_{43} & t_{44}
  \end{array}
\right)
\left(
  \begin{array}{cccc}
    1 &  0 &  0 & -i  \\
    0 & -i & -1 &  0  \\
    0 & -i &  1 &  0  \\
    1 &  0 &  0 &  i 
  \end{array}
\right)  \\
&=&\frac{1}{2}
\left(
  \begin{array}{cccc}
    t_{11}+t_{41}+t_{14}+t_{44} & -i(t_{12}+t_{42}+t_{13}+t_{43}) & 
    -t_{12}-t_{42}+t_{13}+t_{43} & i(-t_{11}-t_{41}+t_{14}+t_{44}) \\
    i(t_{21}+t_{31}+t_{24}+t_{34}) & t_{22}+t_{32}+t_{23}+t_{33} & 
    i(-t_{22}-t_{32}+t_{23}+t_{33}) & t_{21}+t_{31}-t_{24}-t_{34} \\
    -t_{21}+t_{31}-t_{24}+t_{34} & i(t_{22}-t_{32}+t_{23}-t_{33}) & 
    t_{22}-t_{32}-t_{23}+t_{33} & i(t_{21}-t_{31}-t_{24}+t_{34}) \\
    i(t_{11}-t_{41}+t_{14}-t_{44}) & t_{12}-t_{42}+t_{13}-t_{43} & 
    i(-t_{12}+t_{42}+t_{13}-t_{43}) & t_{11}-t_{41}-t_{14}+t_{44}    
  \end{array}
\right).
\end{eqnarray*}
\end{footnotesize}
\vspace{-5mm}
\begin{equation}
{}
\end{equation}

Last, we calculate each component above by making use of 
(\ref{eq:Second}).  If we set
\begin{equation}
e^{A}=R^{\dagger}TR=X=(x_{ij})
\end{equation}
then straightforward but long algebra gives
\begin{eqnarray*}
x_{11}
&=&\frac{1}{2}(t_{11}+t_{41}+t_{14}+t_{44})  \\
&=&\cos|a|\cos|b|-\frac{\sin|a|\sin|b|}{|a||b|}(a_{1}b_{1}-a_{2}b_{2}+a_{3}b_{3}), \\
x_{21}
&=&\frac{i}{2}(t_{21}+t_{31}+t_{24}+t_{34})  \\
&=&-\cos|a|\frac{\sin|b|}{|b|}b_{1}-\frac{\sin|a|}{|a|}\cos|b|a_{1}
+\frac{\sin|a|\sin|b|}{|a||b|}(a_{2}b_{3}+a_{3}b_{2}), \\
x_{31}
&=&\frac{1}{2}(-t_{21}+t_{31}-t_{24}+t_{34})  \\
&=&\cos|a|\frac{\sin|b|}{|b|}b_{2}-\frac{\sin|a|}{|a|}\cos|b|a_{2}
-\frac{\sin|a|\sin|b|}{|a||b|}(a_{1}b_{3}-a_{3}b_{1}), \\
x_{41}
&=&\frac{i}{2}(t_{11}-t_{41}+t_{14}-t_{44})  \\
&=&-\cos|a|\frac{\sin|b|}{|b|}b_{3}-\frac{\sin|a|}{|a|}\cos|b|a_{3}
-\frac{\sin|a|\sin|b|}{|a||b|}(a_{1}b_{2}+a_{2}b_{1}); \\
\end{eqnarray*}
\begin{eqnarray*}
x_{12}
&=&\frac{-i}{2}(t_{12}+t_{42}+t_{13}+t_{43})  \\
&=&\cos|a|\frac{\sin|b|}{|b|}b_{1}+\frac{\sin|a|}{|a|}\cos|b|a_{1}
+\frac{\sin|a|\sin|b|}{|a||b|}(a_{2}b_{3}+a_{3}b_{2}), \\
x_{22}
&=&\frac{1}{2}(t_{22}+t_{32}+t_{23}+t_{33})  \\
&=&\cos|a|\cos|b|-\frac{\sin|a|\sin|b|}{|a||b|}(a_{1}b_{1}+a_{2}b_{2}-a_{3}b_{3}), \\
x_{32}
&=&\frac{i}{2}(t_{22}-t_{32}+t_{23}-t_{33})  \\
&=&\cos|a|\frac{\sin|b|}{|b|}b_{3}-\frac{\sin|a|}{|a|}\cos|b|a_{3}
+\frac{\sin|a|\sin|b|}{|a||b|}(a_{1}b_{2}-a_{2}b_{1}), \\
x_{42}
&=&\frac{1}{2}(t_{12}-t_{42}+t_{13}-t_{43})  \\
&=&\cos|a|\frac{\sin|b|}{|b|}b_{2}+\frac{\sin|a|}{|a|}\cos|b|a_{2}
-\frac{\sin|a|\sin|b|}{|a||b|}(a_{1}b_{3}+a_{3}b_{1}); \\
\end{eqnarray*}
\begin{eqnarray*}
x_{13}
&=&\frac{1}{2}(-t_{12}-t_{42}+t_{13}+t_{43})  \\
&=&-\cos|a|\frac{\sin|b|}{|b|}b_{2}+\frac{\sin|a|}{|a|}\cos|b|a_{2}
-\frac{\sin|a|\sin|b|}{|a||b|}(a_{1}b_{3}-a_{3}b_{1}), \\
x_{23}
&=&\frac{i}{2}(-t_{22}-t_{32}+t_{23}+t_{33})  \\
&=&-\cos|a|\frac{\sin|b|}{|b|}b_{3}+\frac{\sin|a|}{|a|}\cos|b|a_{3}
+\frac{\sin|a|\sin|b|}{|a||b|}(a_{1}b_{2}-a_{2}b_{1}), \\
x_{33}
&=&\frac{1}{2}(t_{22}-t_{32}-t_{23}+t_{33})  \\
&=&\cos|a|\cos|b|+\frac{\sin|a|\sin|b|}{|a||b|}(a_{1}b_{1}+a_{2}b_{2}+a_{3}b_{3}), \\
x_{43}
&=&\frac{i}{2}(-t_{12}+t_{42}+t_{13}-t_{43})  \\
&=&\cos|a|\frac{\sin|b|}{|b|}b_{1}-\frac{\sin|a|}{|a|}\cos|b|a_{1}
-\frac{\sin|a|\sin|b|}{|a||b|}(a_{2}b_{3}-a_{3}b_{2}); \\
\end{eqnarray*}
\begin{eqnarray*}
x_{14}
&=&\frac{i}{2}(-t_{11}-t_{41}+t_{14}+t_{44})  \\
&=&\cos|a|\frac{\sin|b|}{|b|}b_{3}+\frac{\sin|a|}{|a|}\cos|b|a_{3}
-\frac{\sin|a|\sin|b|}{|a||b|}(a_{1}b_{2}+a_{2}b_{1}), \\
x_{24}
&=&\frac{1}{2}(t_{21}+t_{31}-t_{24}-t_{34})  \\
&=&-\cos|a|\frac{\sin|b|}{|b|}b_{2}-\frac{\sin|a|}{|a|}\cos|b|a_{2}
-\frac{\sin|a|\sin|b|}{|a||b|}(a_{1}b_{3}+a_{3}b_{1}), \\
x_{34}
&=&\frac{i}{2}(t_{21}-t_{31}-t_{24}+t_{34})  \\
&=&-\cos|a|\frac{\sin|b|}{|b|}b_{1}+\frac{\sin|a|}{|a|}\cos|b|a_{1}
-\frac{\sin|a|\sin|b|}{|a||b|}(a_{2}b_{3}-a_{3}b_{2}), \\
x_{44}
&=&\frac{1}{2}(t_{11}-t_{41}-t_{14}+t_{44})  \\
&=&\cos|a|\cos|b|+\frac{\sin|a|\sin|b|}{|a||b|}(a_{1}b_{1}-a_{2}b_{2}-a_{3}b_{3}).
\end{eqnarray*}

This completes the calculation of $e^{A}\ (A\in so(4))$. The form is 
of course not simple, while it is not so ugly in the Dirac sense.  
It is straightforward to verify the orthogonality $X^t X=X X^t=I_4$ 
and the unit determinant det\,$X=1$.
See \cite{RZ}, \cite{FO}, \cite{F} for related topics.

\vspace{3mm}\noindent
{\bf Comment}\ \ 
The matrix $R$ is called the magic one by Makhlin. Readers must 
understand the reason why it is called {\bf magic} through this paper.

\newpage
Is it possible  to calculate $e^{A}$ for $A\in so(n)\ (n\geq 5)$ ? 
Unfortunately or furtunately, it has not been done at the present time, so we leave 
the problem to young readers.

\vspace{3mm}
Last, let us present an exercise to readers. 

\noindent
{\bf Exercise}\ \ The three--dimensional special orthogonal group 
$SO(3)\ (\cong SU(2)/{\bf Z}_{2})$ is a subgroup of $SO(4)$ and 
can be embedded into $SO(4)$ like
\[
SO(3)\ \longrightarrow \ SO(4)\ :\ 
O\ \longmapsto \ 
\left(
  \begin{array}{cc}
    O &   \\
       & 1
  \end{array}
\right).
\]
Therefore, for $B\in so(3)$
\begin{equation}
B=
\left(
  \begin{array}{ccc}
    0   & a  & c  \\
    -a & 0  & b  \\
    -c & -b & 0
  \end{array}
\right)
\end{equation}
we can write
\begin{equation}
\label{eq:reduction}
\left(
  \begin{array}{cc}
    e^{B} &   \\
           & 1
  \end{array}
\right)
=
\exp
\left\{
\left(
  \begin{array}{cc}
    B &   \\
       & 0
  \end{array}
\right)
\right\}
=
\exp
\left\{
\left(
  \begin{array}{cccc}
    0   & a  & c & 0 \\
    -a & 0  & b & 0 \\
    -c & -b & 0 & 0 \\
    0  & 0   & 0 & 0
  \end{array}
\right)
\right\}.
\end{equation}
Since 
\[
\left(
  \begin{array}{cccc}
    0   & a  & c & 0 \\
    -a & 0  & b & 0 \\
    -c & -b & 0 & 0 \\
    0  & 0   & 0 & 0
  \end{array}
\right)\in so(4)
\]
we can easily calculate (\ref{eq:reduction}) and obtain $e^{B}$ 
from the result in the preceding section.  

\noindent
Carry out this procedure.

\section{Concluding Remarks}

In this paper we wrote down $\exp A\ (A\in so(4))$ explicitly. 
Its form is relatively simple and beautiful, and has not been given 
as far as we know.

While preparing this paper, a very sad news of  Steve Jobs' death arrived. 
With grief I would like to dedicate this paper  to his memory. 
His way of thinking is in my opinion based on three words
\begin{center}
(1)\ simple,\quad (2)\ easy to use,\quad (3)\ beautiful.
\end{center}
These may be unified as ``super smart". It must be very important 
in almost all fields. 
Whether the result in the paper is  ``super smart" or not will 
be left to readers.

He also says
\begin{center}
Stay hungry,\quad Stay foolish.
\end{center}
That must be the spirit of Apple or Silicon Valley, and the spirit is 
strongly required by not only USA but also Japan.

\vspace{3mm}
We conclude this paper with the recent hot topic $\cdots$ the OPERA 
experiment results \cite{OPERA}.  Namely, {\bf the speed of neutrino 
may exceed that of light in vacuum}. The result is ``foolish"  
enough.  At the present time 
it is impossible to conclude the superluminal neutrinos, so we are 
looking forward to the ``super smart" interpretation of \cite{OPERA}. 
See also \cite{MB} and its references for a possibility.

\vspace{5mm}
The author would like to thank Ryu Sasaki for useful comments and 
suggestions.

%%%%%%%%%%%%%
%References%
%%%%%%%%%%%%%

\end{document}